\documentclass[journal=jacsat,manuscript=article,layout=twocolumn]{achemso}

\usepackage{chemformula} 
\usepackage[T1]{fontenc} 

\usepackage{hyperref}
\usepackage{siunitx}
\usepackage[version=3]{mhchem}
\usepackage{gensymb}
\usepackage{float}

\usepackage{cuted}

\author{Dana O. Byrne}
\affiliation[1]
{Department of Chemistry, University of California, Berkeley, CA, USA}
\alsoaffiliation[2]
{Department of Materials Science and Engineering, University of California, Berkeley, CA, USA}
\alsoaffiliation[3]
{National Center for Electron Microscopy, Molecular Foundry, Lawrence Berkeley National Laboratory, CA, USA}
\email{dana_byrne@berkeley.edu}
\author{Frances I. Allen}
\email{francesallen@berkeley.edu}
\affiliation[2]
{Department of Materials Science and Engineering, University of California, Berkeley, CA, USA}
\alsoaffiliation[3]
{National Center for Electron Microscopy, Molecular Foundry, Lawrence Berkeley National Laboratory, CA, USA}
\alsoaffiliation[4]
{California Institute for Quantitative Biosciences, University of California, Berkeley, CA, USA}

\title[Precision Engineering of Triangular Nanopores in Monolayer hBN]
  {Atomic Engineering of Triangular Nanopores in Monolayer hBN: A Decoupled Seeding and Growth Approach}


\begin{document}








\begin{strip}
\centering
\begin{minipage}{0.85\textwidth}

\begin{abstract}

Nanopores in 2D materials are of significant interest in advanced membrane technologies aimed at the sensing and separation of ions and molecules. These applications necessitate 2D nanopores that are precise in size and shape, and abundant in number. However, conventional fabrication techniques often struggle to achieve both high precision and throughput. In this study, we introduce a decoupled seeding and growth approach designed to overcome this limitation. The method allows the controlled fabrication of ensembles of nanopores with narrow size distribution and is demonstrated for free-standing monolayer hexagonal boron nitride. Using light ion showering, we first create vacancy defect seeds. These seeds are then expanded into triangular nanopores through element-specific preferential atom removal under broad-beam electron irradiation in a transmission electron microscope. Nanopore density and size are controlled by the ion and electron irradiation doses, respectively. During the electron irradiation step, high-resolution imaging allows real-time tracking of the nanopore formation process, enabling the highest level of control. An additional workflow is introduced using thermal annealing in air for the nanopore growth step, delivering the most flexible platform for nanopore fabrication over larger areas with the significant benefit of concurrently removing surface hydrocarbon contamination to mitigate pore clogging and distortion. This study provides researchers with a novel approach to create ensembles of meticulously designed nanopores with well-controlled size and geometry, facilitating the development of next-generation membrane devices that will demand high precision and high throughput nanopore fabrication pipelines.\\

\end{abstract}

\end{minipage}
\end{strip}

\newpage
\section{Introduction}

Nanoporous 2D materials are increasingly leveraged for advanced filtration, separation, and sensing applications, 
including ion sieving, water desalination~\citep{Fu2020Dehydration-DeterminedSubnanopores,Surwade2015WaterGraphene}, and the detection of single biomolecules~\citep{Garaj2013Molecule-huggingNanopores,Garaj2010GrapheneMembrane}.
The advantageous properties of 2D materials for these applications include their atomic thinness, which minimizes transport resistance and hence enhances permeability, as well as their high selectivity, owing to tunable pore sizes and opportunities for chemical functionalization~\citep{Su2021}. For the selective transport of ions and small molecules, precisely sized sub-nanometer pores are essential~\citep{Fang2019}.
However, to achieve efficient transport, these precise sub-nanometer pores should be distributed at high density over large areas. Fabricating nanopores to meet these stringent requirements is very challenging, since atomic-level control typically compromises throughput. Conversely, methods that prioritize high throughout often lack precise control over nanopore size and shape.

Thus, there is a critical need for nanopore fabrication methods that can achieve both precise nanopore geometry and high throughout to enable the production of meticulously designed 2D nanopore ensembles for practical applications.

Conventional nanopore fabrication in 2D materials typically employs top-down approaches such as focused ion/electron beam milling~\citep{Macha2022a,Gilbert2017}, plasma treatment~\citep{Surwade2015WaterGraphene}, chemical etching~\citep{Thiruraman2020IonsZero}, and dielectric breakdown~\citep{Guo2022}. Each method has its distinct advantages and drawbacks~\citep{Su2021} . 
Focused ion and electron beam milling techniques are known for the highest level of precision in both positioning accuracy and control over nanopore size and shape.
For example, focused ion beam (FIB) milling using helium ions, with a nominal probe size of 0.5 nm, has been used to fabricate pores approaching 1 nm in diameter in free-standing 2D materials~\citep{Allen2021} by dwelling the probe onto a single point and sputtering atoms from the backside via forward momentum transfer (so-called knock-on). Taking this method to its extreme, focused electron beam milling, employing a sub-Ångström scanning transmission electron microscope (STEM) probe, can achieve the targeted manipulation of individual atoms through a combination of elastic and inelastic scattering effects~\citep{Susi2019}. However, a drawback of FIB and STEM processing is their serial nature due to the single focused probe, resulting in low throughput.

As an alternative to serial FIB milling, several groups have demonstrated the use of FIB microscopes in ``shower mode'', producing multiple sub-nanometer pores simultaneously in graphene and 2D \ce{MoS2}~\citep{Russo2012, Yoon2016b, Macha2022a}. In this mode, the ion beam is rastered at low dose over broader areas, such that only a few ions impact the sample per square nanometer.
Some of these collisions transfer momentum directly to the atoms in the 2D lattice, causing their removal. This process creates a stochastic areal distribution of single to multi-atom vacancy defects arising from single ion hits.  
By tuning the dose, the ion beam shower method offers substantial control over the density of vacancy defects and is relatively high throughput, capable of irradiating tens to hundreds of square micrometers within minutes. 
However, the FIB shower method offers limited control over the size and shape of the vacancy defects/nanopores that are generated.

\begin{figure*}
 \centering
 \includegraphics[width=0.9\textwidth]{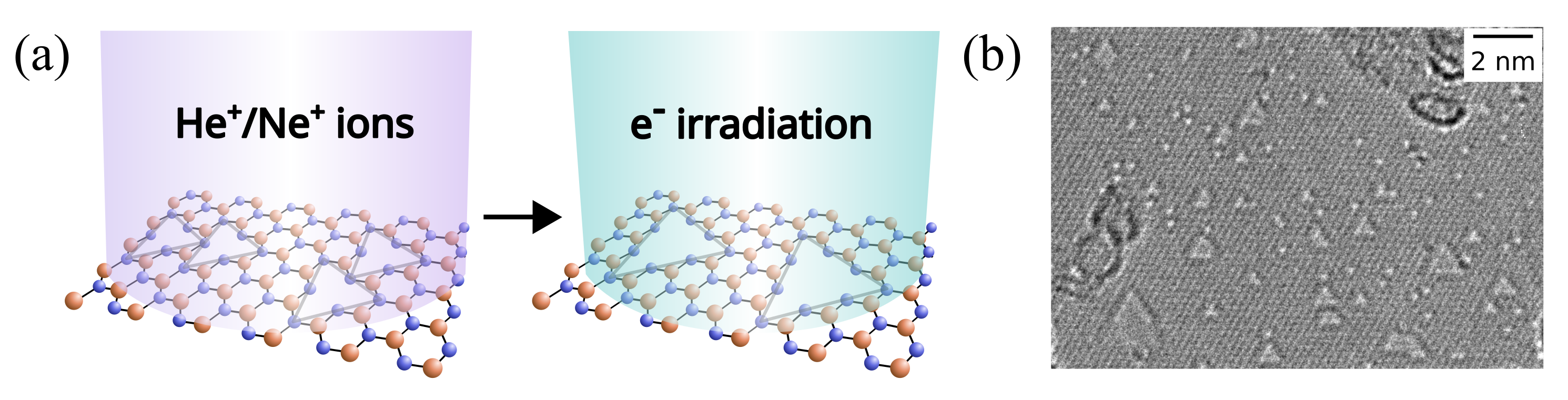}
 \caption{(a) Schematic showing He/Ne ion irradiation to seed vacancy defects in free-standing monolayer hBN and subsequent electron irradiation to expand the defects. (b) HR-TEM image of Ne ion seeded, electron beam grown triangular vacancy defects in monolayer hBN.}
 \label{Fig:Fabrication}
\end{figure*}
 
In another approach, geared towards gaining more control over the nanopore dimensions at increased throughput, broad-beam electron irradiation in the transmission electron microscope (TEM) can be employed. This method combines simultaneous nanopore growth due to wide-field electron bombardment with in-situ imaging at high spatial resolution to monitor the process in real time.  
Furthermore, by taking advantage of heteroatomic 2D material structures like hexagonal boron nitride (hBN), nanopores with specific geometries can be obtained. For example, when irradiating 2D hBN's alternating B and N honeycomb structure with \SI{80}{keV} electrons (at room temperature), preferential boron removal occurs leading to the formation of triangular nanopores with nitrogen terminated edges~\citep{Meyer2009,Alem2009,Alem2011VacancyIrradiation,Kotakoski2010,Ryu2015Atomic-scaleIrradiation,Gilbert2017}.
The starting points (or seeds) for the individual nanopores fabricated by this TEM method are either random intrinsic vacancy defects, or they can can be vacancy defects formed by condensing the electron beam to increase the local electron dose rate, which increases the probability to eject single atoms~\citep{Gilbert2017}. However, the throughput of the TEM technique is ultimately limited by beam current, which forces the area of the condensed electron beam on the sample down to \SIrange[range-units=single,range-phrase=--]{10}{100}{nm} in diameter in order to achieve the required seeding dose rates.

To address the various limitations described above, we introduce a decoupled ion and electron irradiation approach, combining the throughput and density control of the FIB shower method~\citep{Macha2022a, Yoon2016b} with the size and geometry control of the broad-beam TEM method~\citep{Gilbert2017, Ryu2015Atomic-scaleIrradiation,Alem2011VacancyIrradiation} to reliably produce distributions of nanopores of desired shape and narrow size distribution. Specifically, we first shower free-standing hBN monolayers with \SI{25}{keV} He or Ne ions to produce high-density distributions of vacancy defects as determined by the ion dose, and then controllably expand these vacancy defects using \SI{80}{keV} electrons under broad beam illumination in the TEM with direct imaging at atomic resolution in-situ (Fig.~\ref{Fig:Fabrication}(a)). The end result is a high density of triangular nanopores within a tight size range (Fig.~\ref{Fig:Fabrication}(b)). The narrow size distribution of the final nanopore ensemble is critically enabled by producing the initial single-atom vacancy defect seeds using the light-ion beam shower method.

The hBN nanopore fabrication approach demonstrated here opens up new possibilities to prototype novel ion transport concepts that require large numbers of nanopores for efficiency, together with precise nanopore geometry and electrostatic environment for selectivity~\citep{Smolyanitsky2018,Fang2019,Smolyanitsky2020,Violet2024}. With such applications in mind we also introduce an adapted workflow, which after the ion seeding step, uses thermal treatment in air to grow the nanopores and concurrently remove surface hydrocarbon contamination that otherwise causes pore distortion and clogging. Thus, a simple method to create high-density triangular nanopores of narrow size distribution over large contamination-free areas is realized, which can now be leveraged to innovate ion transport technologies. 

\section{Results and Discussion}

\begin{figure}[t]
 \centering
 \includegraphics[height = 9.5cm]{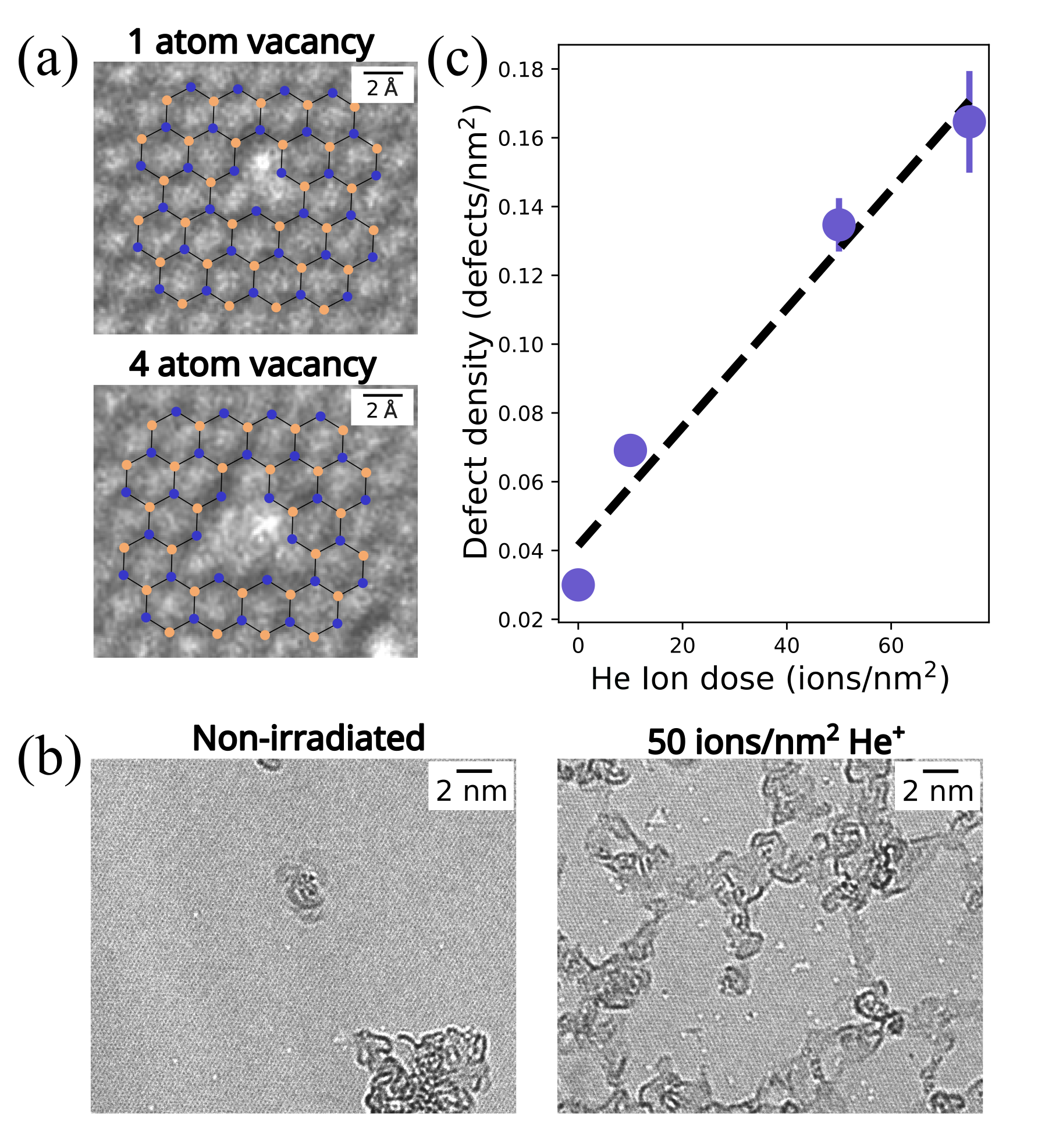}
 \caption{(a) HR-TEM examples of a single atom (top) and a four atom (bottom) N-terminated vacancy defect fabricated by He ion irradiation. The model overlay shows N atoms in blue and B atoms in orange. (b) HR-TEM comparison of monolayer hBN with no ion irradiation vs.\ 50 ions/nm$^2$ He ion irradiation. (c) Scatter plot with best fit line indicating linear relationship between ion irradiation dose and defect density.}
 \label{Fig:Ion seeds}
\end{figure}

\subsection{Ion seeding}

As outlined in Fig.~\ref{Fig:Fabrication}, we first use irradiation with light ions (He$^+$ or Ne$^+$) to generate the starting vacancy defects for our desired ensemble of nanopores in the monolayer hBN. These point defects are generated via knock-on damage, which refers to the elastic transfer of energy from the accelerated ions to the B and N atoms. Momentum transfer from the ions to the atoms causes the atoms to sputter in the forward direction, resulting in single-atom to small multi-atom (up to four atom) vacancies in the hBN, see Fig.~\ref{Fig:Ion seeds}(a). The ion dose is tuned such that the collisions do not `pile up', i.e.\ the defects are formed due to single ion hits. Therefore, the defect seeds produced by the showering He and Ne ions have a very narrow size distribution, with a density determined by the ion dose. Since the defect seeds have similar starting sizes, they act as consistently sized starting points for the following nanopore growth step. 

The vacancy defects produced by the ion irradiation step can be either B or N vacancies, since for 25 keV He/Ne ions, the knock-on interaction transfers enough momentum to well exceed the threshold of ejection for either atom (\SI{\sim20}{eV}, slightly higher for B versus N).~\citep{Kotakoski2010} In the case of the one-atom and four-atom vacancy defects shown in Fig.~\ref{Fig:Ion seeds}(a), these were determined to be N terminated (as shown by the model overlay). This is because the orientation direction of these triangular defects matches that of the TEM-grown expanded triangular defects subsequently grown for the same field of view, which by the TEM method at \SI{80}{keV} have been shown to be N terminated~\citep{Meyer2009,Alem2009}. (TEM growth results are shown in the following section.)

 Figure~\ref{Fig:Ion seeds}(b) shows representative TEM images of monolayer hBN that (on the left) was not showered with ions and (on the right) was showered with \SI{50}{ions/nm\textsuperscript{2}} He ions. Although the non-irradiated hBN monolayer (grown by chemical vapor deposition (CVD)) has some native vacancy point defects, the He ion irradiated monolayer has a much higher overall defect density of approximately \SI{0.13}{defects/nm\textsuperscript{2}}, some of which are four-atom vacancies. A statistical analysis of defect density for a series of irradiated samples shows a linear increase in defect density with increasing ion irradiation dose (Fig.~\ref{Fig:Ion seeds}(c)), demonstrating the fine control over defect density achieved using this ion seeding method. Defect distributions for a given dose were fairly uniform over the irradiated areas, which can be up to hundreds of square micrometers in size. 
 
 The ion irradiation procedure also causes an increase in the level of hydrocarbon contamination on the hBN monolayer samples as the ion dose increases. This can be seen in Fig.~\ref{Fig:Ion seeds}(b), where the \SI{50}{ions/nm\textsuperscript{2}} irradiated hBN monolayer has significantly more hydrocarbon coverage than the non-irradiated sample. This phenomenon is due to the mobilization and decomposition of residual hydrocarbons present on the sample and in the sample chamber under the scanning ion beam~\citep{Hlawacek2013ToSEM}. 
 Mitigating hydrocarbon formation in each step is important for reducing surface contamination on the final sample to facilitate imaging and to help avoid pore clogging in ion/molecule transport experiments.

\subsection{TEM growth}

Once the vacancy defect seeds are produced, the hBN is transferred to the TEM for the controlled growth of the vacancy defects into larger, triangular nanopores under \SI{80}{keV} electron irradiation. Growth dose rates ranged from 800--2000 e/\AA\textsuperscript{2}/s (8$\times$10$^4$--2$\times$10$^5$ e/nm\textsuperscript{2}/s) and growth times ranged from 1-20 minutes, depending on the desired final pore size. Elastic knock-on collisions with preferential B ejection at an electron beam energy of \SI{80}{keV} are generally thought to be responsible for the defects' consistent triangular shape and nitrogen-terminated edges, with inelastic contributions in the form of charging of the insulating hBN likely also playing a role~\citep{Cretu2015,Susi2019}. Because this growth step is performed in the TEM, high-resolution (HR) TEM imaging can be used for real-time monitoring of the pore growth process, and the growth can then be stopped when the nanopores reach a desired size. Nanopore growth using electron irradiation in the TEM is therefore a highly controlled method, and when preceded by the ion beam shower, can produce thousands of triangular nanopores with narrow size distribution. 

\begin{figure}[h]
\centering
  \includegraphics[height = 10 cm]{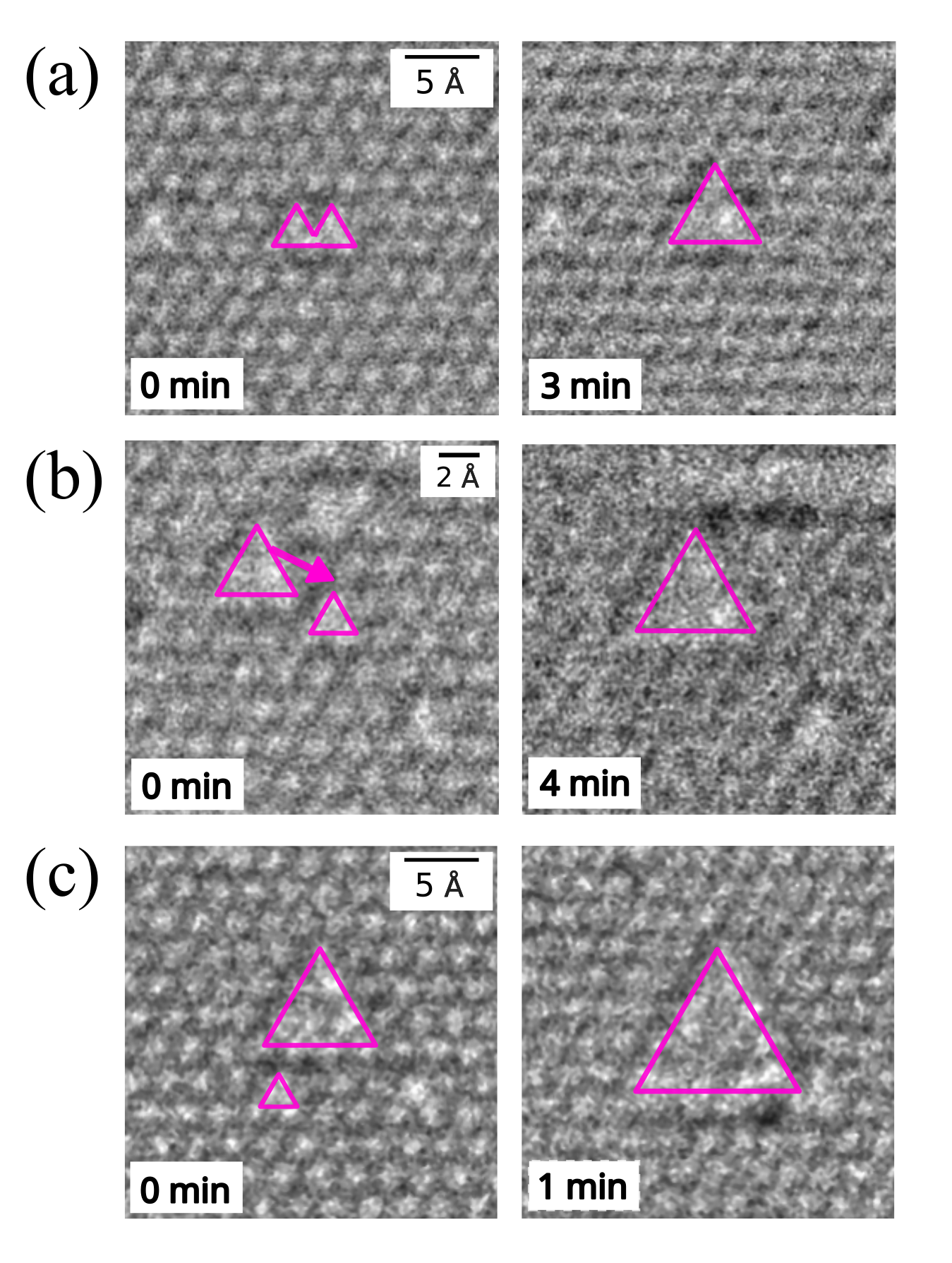}
  \caption{HR-TEM images acquired before and after electron beam-induced merging and growth starting from (a) two single atom vacancies, (b) a four atom and a single atom vacancy, and (c) a nine atom and a single atom vacancy. The growth dose rate for all three sets of images was 1500 e/\AA\textsuperscript{2}/s.}
  \label{Fig:Defect merging}
\end{figure}

Various nanopore growth mechanisms were observed, including single nanopore expansion\citep{Gilbert2017}, nanopore merging~\citep{Ryu2015Atomic-scaleIrradiation}, and defect hopping\citep{Alem2011VacancyIrradiation}. Fig.~\ref{Fig:Defect merging} shows three different types of nanopore growth observed in the same 100~ions/nm\textsuperscript{2} He ion irradiated hBN monolayer sample during electron irradiation at a dose rate of 1500 e/\AA\textsuperscript{2}/s. In Fig.~\ref{Fig:Defect merging}(a), two first neighbor single B vacancies combine to form one four-atom vacancy via the removal of an N atom and an additional B atom, a mechanism representing single nanopore expansion\citep{Gilbert2017}. Another mechanism of defect growth that is demonstrated in Fig.~\ref{Fig:Defect merging}(b) and in other reports,\citep{Alem2011VacancyIrradiation} sees two second neighbor B vacancies merging as a result of one vacancy migrating towards the other before combining. Fig.~\ref{Fig:Defect merging}(c) shows a nine-atom vacancy that has grown towards a single B vacancy before the two vacancies combine to form one larger vacancy. This mechanism was most often observed in the case of two neighboring larger vacancies, which then merged together, or when one larger expanding vacancy mergers with other smaller vacancy defects. 

The growth mechanisms listed here are not exhaustive, with several other electron beam-induced hBN defect growth mechanisms having been reported in the literature.\citep{Alem2011VacancyIrradiation, Gilbert2017} The full image series capturing the defects highlighted in Fig.~\ref{Fig:Defect merging} is shown in the Supplementary Information Fig.~S1. 

The vacancy defects seeded by the He and Ne ions are generally similar in size (mainly single-atom vacancies), but the higher sputter yield ($\approx$11\% for Ne vs.\ $\approx$0.3\% for He, calculated from the defect densities in the respective ion-irradiated hBN samples) means that lower Ne doses can be used to achieve a given defect density. We also observe higher ratios of closely spaced vacancies in the Ne ion irradiated samples, i.e.\ groups of three or more single atom vacancies separated by only one to three atoms. These clusters are observed to combine to form larger nanopores with lower applied electron dose as compared to isolated single defects, which require a higher electron dose to grow. The relative ease for such clusters to grow into larger nanopores is likely due to a lower energy barrier associated with the merging of multiple closely spaced vacancies at the beginning of the growth process.

\begin{figure*}[t]
 \centering
 \includegraphics[width=0.8\textwidth]{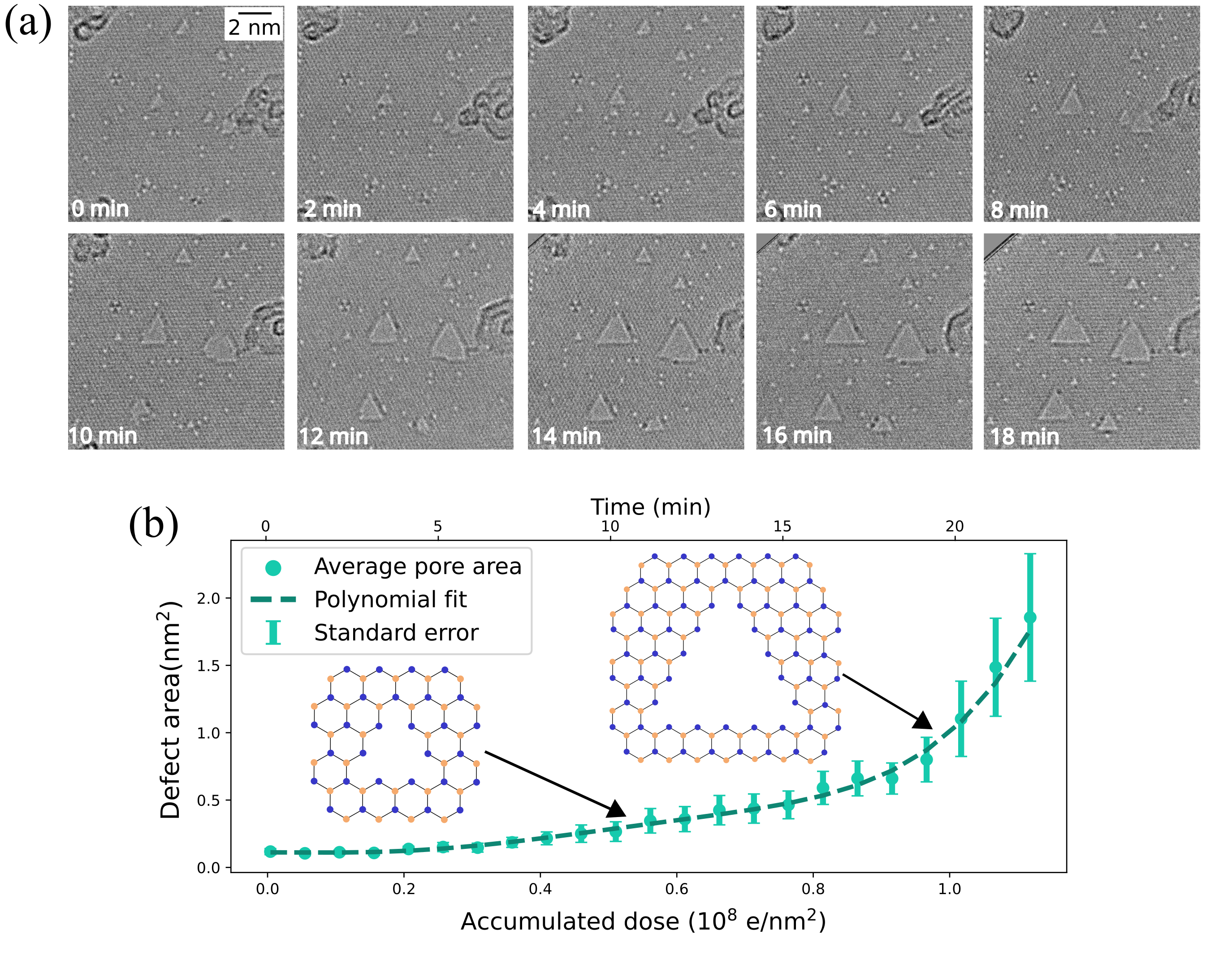}
 \caption{(a) A time series of HR-TEM images taken 2 minutes apart showing Ne ion seeded vacancy defect clusters merging into triangular nanopores as a result of electron irradiation. The electron beam dose rate for growth was 840--870\,e$^-$/\AA$^2$/s. (b) Average defect area growth trajectory with standard errors calculated from the full image series sampled in (a). The inset of (b) shows two atomic models demonstrating the configurations of a 4 atom (left) and a 25 atom (right) nanopore with arrows indicating where these nanopores lie on the growth trajectory plot.}
 \label{Fig:growth series}
\end{figure*}

Figure~\ref{Fig:growth series}(a) shows a nanopore growth time series for monolayer hBN that had been irradiated with 1 ion/nm\textsuperscript{2} Ne to achieve a high starting defect density. A video showing the nanopore growth over the full field of view is provided in the Supplementary Information. In this time series, we observe the growth of multi-atom vacancies as well as the merging of small closely spaced vacancies over the course of 18 minutes. Multiple growth mechanisms are seen to occur, including single nanopore expansion and vacancy merging. Closely spaced single atom vacancies appear to spontaneously merge and form larger (>9 atom) defects that start with non-triangular shape. This effect can be seen in Fig.~\ref{Fig:growth series}(a) between 8 and 12 minutes forming the bottom two nanopores (highlighted in further detail in the Supplementary Information Fig.~S2). Irrespective of defect starting size or shape, isolated nanopores always grew towards a triangular geometry oriented in the same direction, as expected due to preferential B atom ejection at \SI{80}{keV}. 

In the growth series sampled in Fig.~\ref{Fig:growth series}(a), 24 different actively growing defects were identified. The polynomial-fitted average vacancy growth trajectory for these 24 defects is shown in Fig.~\ref{Fig:growth series}(b). The plotted vacancy defect areas were determined by manually outlining the defect edges in every other frame (see Supplementary Information video). The magnitudes of the corresponding standard errors indicate that for an accumulated electron dose above \SI{0.9e8}{e/nm\textsuperscript{2}} (corresponding to an average defect area of approx.\ \SI{0.8}{nm\textsuperscript{2}}, which is approximately the size a 25 atom vacancy), the size of the defects at any given time point varied more strongly. In other words, as the average vacancy defect size increases above \SI{1}{nm\textsuperscript{2}}, the size distribution becomes broader. This makes sense, because as the defects grow larger, they are more likely to merge with neighboring defects and thus make drastic jumps in size, as often observed in this study. Isolated larger pores are also expected to grow faster even without merging, due to the lower displacement threshold predicted for vacancies with >9 missing atoms.~\citep{Kotakoski2010} Additionally, surface contaminants were observed to move locally and cause faster pore growth without preference for the triangular shape, presumably due to beam-induced chemical etching~\citep{Jain2024Adatom-mediatedMicroscope} (see Supplementary Information Fig.~S1). 

Not all vacancy defects that were exposed to electron irradiation grew to a larger size. Specifically, single atom vacancies seeded by the ion shower that were spatially isolated from other defects tended to remain the same size, indicating that single atom vacancies can be relatively stable. Single vacancy stability is attributed to the interaction of the respective edge atoms with the next-nearest neighbors in the pristine lattice, giving rise to a larger relative threshold of ejection as compared to the edge atoms of larger defect structures.~\citep{Kotakoski2010} Although single vacancies skew the total size distribution, they are not generally permeable to ions (except protons) and thus should not play a role for ion transport applications like the desalination of water. 

\begin{figure*}[t]
 \centering
 \includegraphics[width=0.95\textwidth]{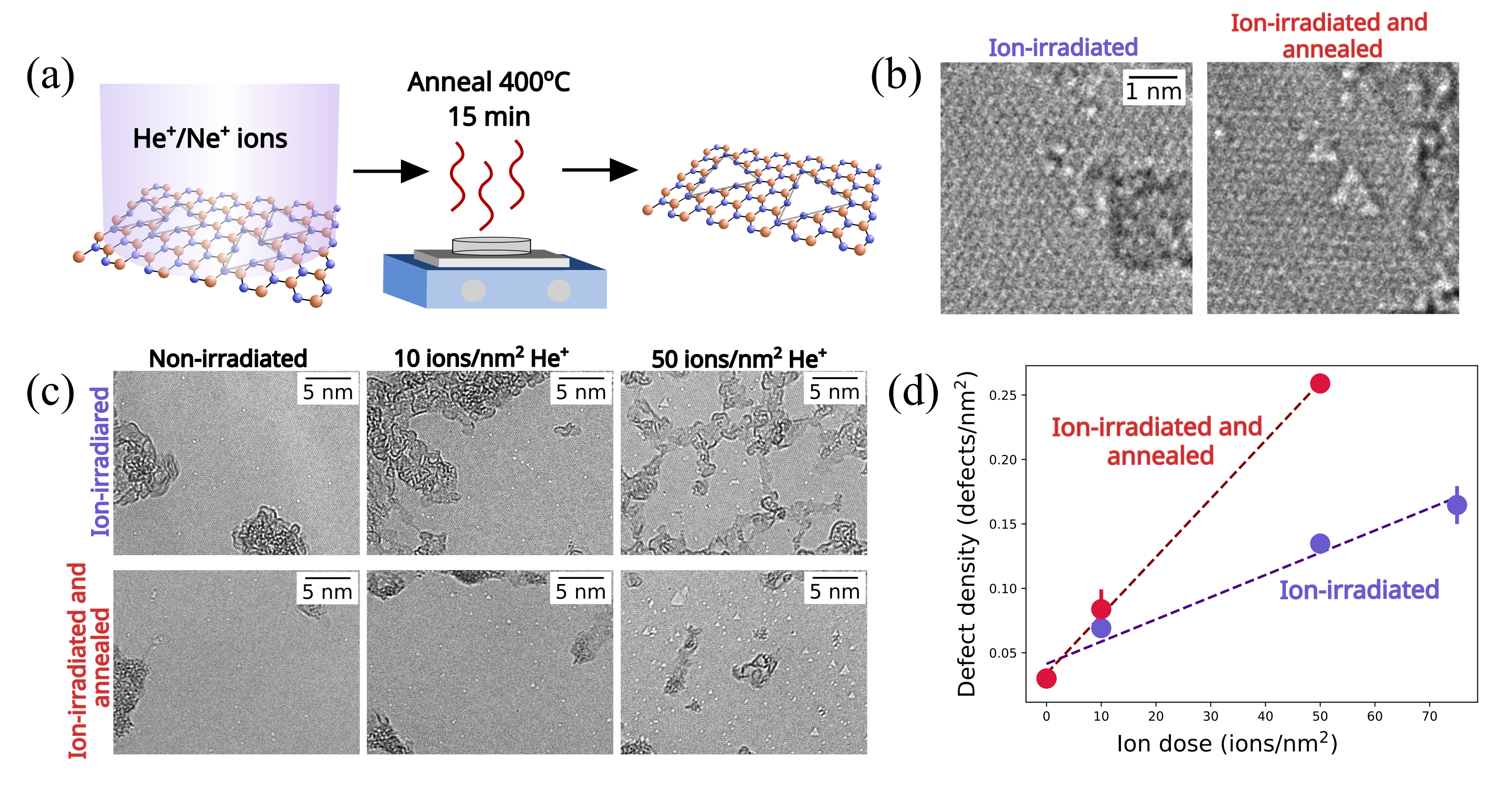}
 \caption{(a) Schematic showing the combined light ion irradiation and subsequent hot plate annealing method. (b) HR-TEM images of 5 ions/nm\textsuperscript{2} Ne ion irradiated monolayer hBN before and after hot plate annealing (same region of sample viewed in each). (c) TEM images of representative monolayer hBN that has undergone no irradiation, 10 ions/nm\textsuperscript{2} He ion irradiation, and 50 ions/nm\textsuperscript{2} He ion irradiation. The second row of images shows representative regions of the same samples after undergoing the hot plate annealing treatment. (d) Scatter plot of defect density vs. He ion irradiation dose with best fit lines.}
 \label{Fig:Hot plate}
\end{figure*}

\subsection{Air annealing}
We have shown that the combined ion shower and electron beam growth method provides control over defect seeding and expansion and significantly improves the throughput of atomically precise nanopore fabrication compared to previous electron-only methods. However, the throughput in this decoupled electron/ion method is ultimately limited by the illumination area of the electron beam (up to 1000 nm\textsuperscript{2} to achieve the required electron dose rate for nanopore growth, constrained by the instrument current). In addition, if samples are covered in hydrocarbon contamination arising from the 2D material transfer or ion-irradiation processes, there may be pore clogging issues during e.g.\ transport experiments. Surface contamination can also affect the defect growth patterns. To address this, we demonstrate an adapted method for high-throughput precision nanopore fabrication that uses the ion-seeding method combined with atmospheric annealing to 1) produce clean large areas of monolayer hBN and 2) grow the triangular nanopores, without the efficiency limitations of the TEM.

Figure~\ref{Fig:Hot plate}(a) shows a schematic of this alternative nanopore fabrication method optimized for throughput and practical application of the nanopores in devices. The hBN monolayers are ion irradiated (as before) but then annealed on a hot plate in ambient air at \SI{400}{\degree C} for 15 minutes. The result of this annealing treatment is monolayer hBN with distributions of expanded triangular nanopores and significantly less surface hydrocarbon contamination. Such a method is not area constrained like the electron growth TEM method, making it a highly scalable technique for fabricating large hBN films with triangular nanopores.

HR-TEM images of the same sample region showing vacancy defects after ion seeding (left) and after the subsequent annealing treatment (right) are shown in Fig.~\ref{Fig:Hot plate}(b). Not all defects grew during the annealing treatment, but the defects that did grow tended towards triangular shapes. Since electron irradiation during the TEM imaging can also alter the pore structure, imaging doses were kept lower than the growth doses by an order of magnitude (4$\times$10$^6$ e/\AA\textsuperscript{2}/s) to minimize this effect. The pore growth mechanism during air annealing presumably involves oxygen chemical etching or a similarly mediated etching process. Inspection of the HR-TEM images of the same sample regions before and after growth suggests that clustered vacancy point defects likely merged to form larger triangles in a similar fashion to the TEM method. Nanopore growth of isolated defects exposed to air without annealing was not observed, indicating that elevated temperatures are necessary for the merging and controlled growth process.

The result that the atmospheric annealing treatment can both remove surface hydrocarbon contamination and grow nanopores at the same time is demonstrated in Fig.~\ref{Fig:Hot plate}(c). Here we see representative TEM images of hBN monolayer without (top row) and with (bottom row) annealing treatment for non-ion-irradiated (left), irradiated with \SI{10}{ions/nm\textsuperscript{2}} He ions (center), and irradiated with \SI{50}{ions/nm\textsuperscript{2}} He ions (right) samples. Lower magnification views are shown in the Supplementary Information Fig.~S3. Two main visual differences are apparent from images of samples with and without annealing: (1) the amount of surface contamination is generally reduced by the anneal and (2) there are more expanded triangular defects after the anneal. On average, the contamination areal coverage is reduced by approximately 60\%, which reveals significantly more of the underlying hBN. The marked increase in average defect size, which is seen for the annealed highest-dose ion irradiated sample, is presumably due to the higher density of starting ion-seeded defects, supporting a merging mechanism of pore growth.

A statistical comparison of defect density vs.\ ion shower dose for ion-irradiated-only and ion-irradiated-and-annealed samples is presented in Fig.~\ref{Fig:Hot plate}(d). Each data point represents the average defect density counted for three distinct sample regions, except for the data point at \SI{75}{ions/nm\textsuperscript{2}}, which corresponds to two distinct sample regions. The standard deviation for each applied ion dose is also plotted for each point. Both data series exhibit a linear trend with ion shower dose, but the ion-irradiated-and-annealed hBN best fit line has a steeper slope. This may be due to a difference in defect density for contamination-free (counted) vs.\ contaminated (uncounted) regions. For example, contamination may preferentially accumulate on high defect density regions, leading to an underestimation of the overall defect density for samples that had more contamination coverage (i.e.\ the non-annealed samples). Contaminant species may also facilitate the formation of new defects during annealing, leading to a relative overestimation of defect density for the annealed samples. In both cases, the result would be a steeper slope in the dose plot for the annealed sample set.  

\section{Conclusions}

In this paper, we demonstrated a decoupled ion and electron irradiation method for fabricating ensembles of triangular nanopores with narrow size distribution in monolayer hBN. Light ion irradiation is first used to seed vacancy defects in the monolayer in controlled densities, after which electron irradiation in the TEM is used to expand these defects into nanopores with triangular shape via preferential B atom removal. The combined ion and electron irradiation approach takes advantage of 1) the throughput and density control of the ion irradiation seeding step, which is performed using a FIB microscope in shower mode, and 2) the shape and size control of electron irradiation step, which is performed in the TEM with real-time monitoring of the nanopores by direct imaging at atomic resolution. 

Multiple mechanisms are observed to contribute to nanopore expansion during the electron irradiation-induced growth stage. These mechanisms predominantly involve various forms of merging of closely-spaced vacancy defect structures. As a result, the narrowest size distribution of the final nanopores is achieved using the lower accumulated electron doses, i.e.\ before exponential pore growth sets in due to the greater propensity for merging once the expanding individual vacancy structures reach a certain spacing. In our experiments, we find that ensembles of nanopores fabricated up to a size of approximately \SI{0.25}{nm\textsuperscript{2}} (corresponding to four-atom vacancies) retain a tight size distribution.
Sub-nanometer pores of this type are of particular interest for e.g.\ the selective transport of Na$^+$ and K$^+$ ions for ion sieving and the desalination of water~\citep{Smolyanitsky2018,Fang2019,Smolyanitsky2020}.

Finally, we also developed an adapted method for triangular nanopore fabrication in monolayer hBN using thermal annealing in air after the ion seeding step to both clean off residual hydrocarbon contamination and grow the nanopores. This integrated cleaning and nanopore growth method has the advantage of providing nanopores that are less likely to become clogged or distorted due to mobile contaminants and which can be fabricated over much larger areas, since one is not limited by the electron beam illumination area. While the precision may be slightly lower, the controlled seeding followed by thermally assisted ex-situ TEM growth method offers researchers a flexible platform to explore ion transport phenomena and prototype new ion transport devices for future practical applications. 

\section{Experimental}

\subsection{Preparation of monolayer hBN}
\subsubsection{Fabrication of custom holey SiN$_x$ TEM supports}
Custom membrane supports were fabricated in house starting from \SI{3}{mm} silicon chips with \SI{20}-{30} {nm}-thick free-standing SiN$_x$ membranes (window size \SI{20}{\micro\m} or \SI{50}{\micro\m}) obtained from Norcada Inc. Into these SiN$_x$ membranes, \SI{200}{nm}-diameter holes were patterned in arrays with index markers using gallium FIB milling (\SI{50}{\pA}, \SI{30}{keV} Ga ions from a Zeiss ORION NanoFab He-Ne-Ga FIB microscope).

\subsubsection{Electrochemical delamination of monolayer hBN}
Suspended monolayer hBN was prepared by electrochemical delamination \cite{Wang2011} of CVD monolayer on copper obtained from Grolltex Inc. For this process, the hBN-on-copper was spin coated with polymethyl methacrylate (PMMA) and clipped into to a two-electrode system with NaCl electrolyte and a glassy carbon cathode. Applying a \SI{5}{V} bias to the system activates the hydrogen evolution reaction, generating hydrogen bubbles at the anode which then remove the hBN/PMMA from the copper surface. The hBN/PMMA films were then washed with de-ionized water and transferred to the custom SiN$_x$ TEM supports before being cleaned with an acetone drip. The acetone drip method was carried out by suspending the sample over acetone that was heated to \SI{150}{\degree C} in an Erlenmeyer flask with a loose covering to prevent excess loss of acetone. Heating the acetone to \SI{150}{\degree C}  causes the solvent to vaporize, condense on the sample, and then drip back into the solution, removing the PMMA from the hBN. Samples were stored under vacuum after fabrication to limit accumulation of surface contamination and uncontrolled defect growth.

\subsection{Vacancy defect seeding with light ions}
To seed vacancy defects and defect clusters in the free-standing hBN monolayers, the samples on the custom holey substrates were irradiated with He or Ne ions using the multibeam He-Ne-Ga FIB operating in shower mode (a low-dose raster) at \SI{25}{keV} (standard beam energy for this instrument). Various target ion doses were used to control defect density: 10, 50, and 75 ions/nm$^2$ for \SI{25}{keV} He ions, and 1 and 5 ions/nm$^2$ for \SI{25}{keV} Ne ions. Scan parameters of \SI{1}{\micro\s} dwell time, variable scan spacing, and variable numbers of repeat scans were used to enable fine control over the spatial range and location of the irradiations, allowing for a single sample to be irradiated site selectively with different ions and various doses, as shown previously~\citep{Byrne2024}. 

\subsection{Electron beam-induced defect growth and imaging}
In order to grow the ion-seeded vacancies in the monolayer hBN into triangular nanopores, samples were transferred into  a double-aberration-corrected modified FEI Titan 80-300 microscope (the TEAM I instrument at the Molecular Foundry, LBNL) equipped with a Continuum K3 direct electron detector. The microscope was operated at \SI{80}{kV} accelerating voltage. Low-dose high-resolution imaging conditions ranged from \SI{500}-{700} {e/\AA\textsuperscript{2}/s}, while defect merging and growth were generally induced using dose rates of \SI{800}-{2000} {e/\AA\textsuperscript{2}/s}. The hBN was illuminated over an area of approximately \SI{50}-{200} {nm} in diameter, which varied depending on the applied dose rate and the beam current.

\subsection{Ex-situ annealing for cleaning and defect growth}
The hBN monolayer samples typically have surface hydrocarbon contamination, which can make direct imaging of the lattice difficult and can cause pore clogging issues for ion/molecule transport applications. Using oxygen annealing experiments reported in the literature as a guide~\citep{Garcia2012EffectiveDevices}, we find that heating the hBN samples at \SI{400}{\degree C} in air for 15 minutes on a hot plate can be used to reduce contamination coverage. This annealing procudure was applied to all samples before the ion irradiation step, to reduce the amount of contamination on the hBN surface from the outset. As shown in Fig.~\ref{Fig:Hot plate}, the annealing treatment is also an alternative way to grow triangular nanopores from the ion-induced vacancy defect seeds in the hBN monolayers.

\subsection{Image processing and statistical analysis}
HR-TEM images were obtained using short acquisition times (\SI{1}-{4}{s}) to minimize the effects of sample drift. Individual frames in a given image series were aligned in DigitalMicrograph\textregistered\ software using automatic band pass filtering with subsequent automatic alignment (and manual alignment as required) to obtain the clearest final image stacks. 

 Defect densities were determined from the HR-TEM image stacks. Areas that were suitable for defect counting were first sectioned off (examples of ineligible regions include those obscured by contamination or bi-layer regions). Defects in these regions were then marked manually and counted by hand in Fiji/ImageJ software~\cite{Schindelin2012}. Phase contrast ambiguities and the potential for beam-induced growth, even at low imaging dose, lead to some uncertainty in the identification of defects down to the single-atom level. Therefore, clusters of closely spaced ($\leq$ 3 atoms apart) single-atom defects and larger defects ($\geq$ 4 atom vacancies) were counted as single defects to avoid over-counting. This in turn means that defect densities and sputter yields will generally be underestimated, especially for higher ion doses where clusters were more prevalent. Example images of defect counting can be found in the Supplementary Information in Fig.~S4. Because counting and sectioning was done by hand, it is possible that a systematic counting error was introduced. However, since the same analysis procedure was used throughout, any trends in defect densities vs.\ dose are expected to be preserved. 

\begin{acknowledgement}

This work was funded in part by NSF Award No.\ 2110924. D.O.B. also acknowledges funding from the Department of Defense through the National Defense Science \& Engineering Graduate (NDSEG) Fellowship Program. Work at the Molecular Foundry was supported by the Office of Science, Office of Basic Energy Sciences, of the U.S. Department of Energy under Contract No.\ DE-AC02-05CH11231. Ion irradiation was performed at the qb3-Berkeley Biomolecular Nanotechnology Center. The authors thank Alex Smolyanitsky for helpful discussions.

\end{acknowledgement}

\begin{suppinfo}

\begin{itemize}
  \item Fig.~S1: Low magnification TEM images corresponding to growth mechanisms shown in Fig.\ 3. \item Fig.~S2: Annotated HR-TEM image series corresponding to Fig.\ 4. 
  \item Fig.~S3: Low magnification TEM images showing effect of annealing treatment on non-irradiated and ion-irradiated hBN corresponding to Fig.\ 5. 
  \item Fig.~S4: Low magnification TEM images showing defect counting corresponding to Figs.\ 2 and 5. 
  \item Video showing electron beam-induced nanopore growth. 
\end{itemize}

\end{suppinfo}

\providecommand{\latin}[1]{#1}
\makeatletter
\providecommand{\doi}
  {\begingroup\let\do\@makeother\dospecials
  \catcode`\{=1 \catcode`\}=2 \doi@aux}
\providecommand{\doi@aux}[1]{\endgroup\texttt{#1}}
\makeatother
\providecommand*\mcitethebibliography{\thebibliography}
\csname @ifundefined\endcsname{endmcitethebibliography}  {\let\endmcitethebibliography\endthebibliography}{}

\clearpage
\onecolumn
\section*{Supplemental Information}
\renewcommand\thefigure{S\arabic{figure}}    
\setcounter{figure}{0}
\begin{figure}[H]
 \centering
 \includegraphics[width=0.9\textwidth]{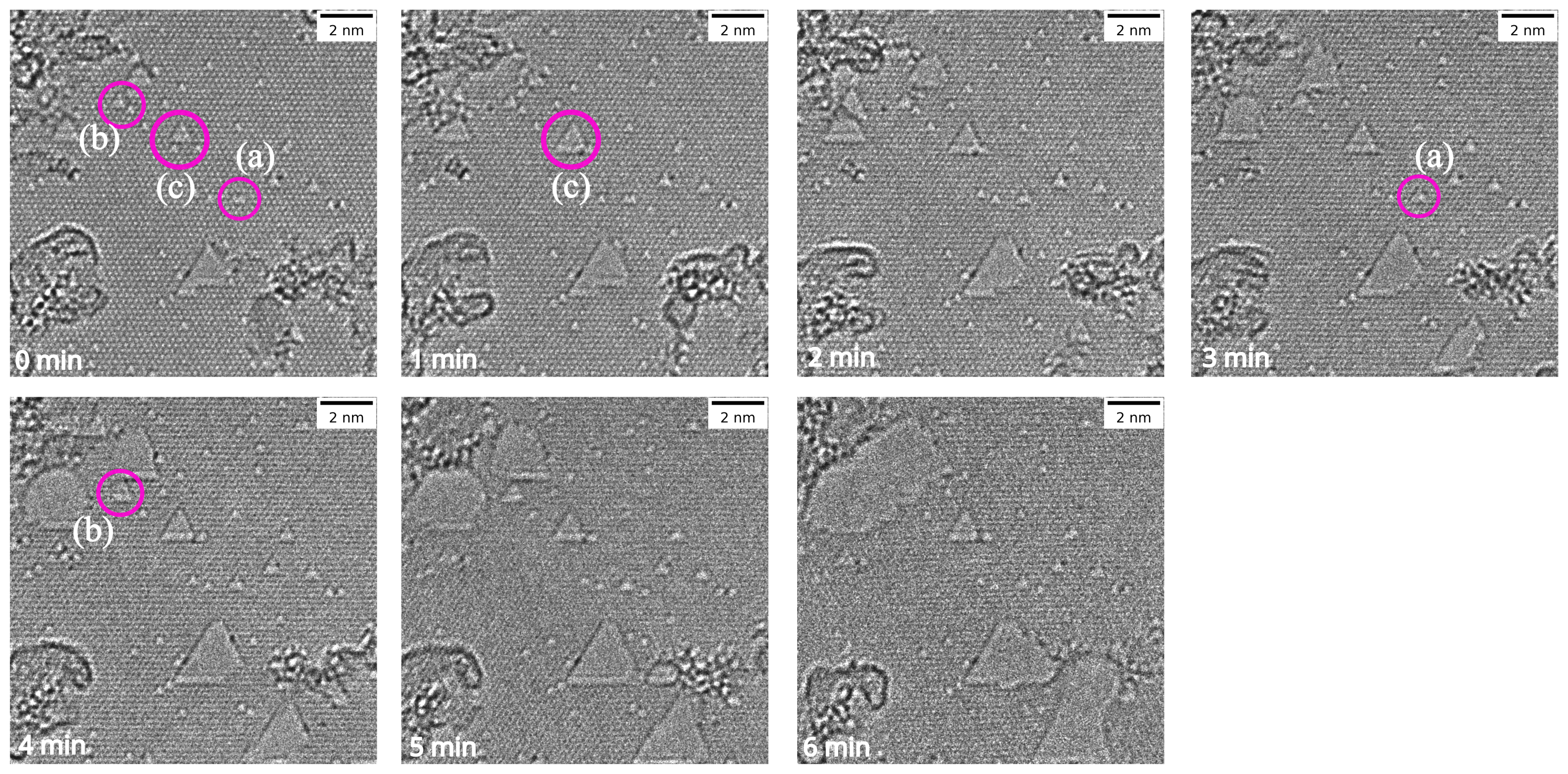}
 \caption{Full TEM image series (zoomed-out) corresponding to the mechanisms shown in Fig.~3 (circles highlight the defects discussed the main text). In the final image, contamination-assisted etching is distinctly observed in the top left and bottom right nanopore regions. The dose rate was 1500 e/\AA\textsuperscript{2}/s.}
 \label{Fig:S1}
\end{figure}

\begin{figure}[H]
 \centering
 \includegraphics[width=0.9\textwidth]{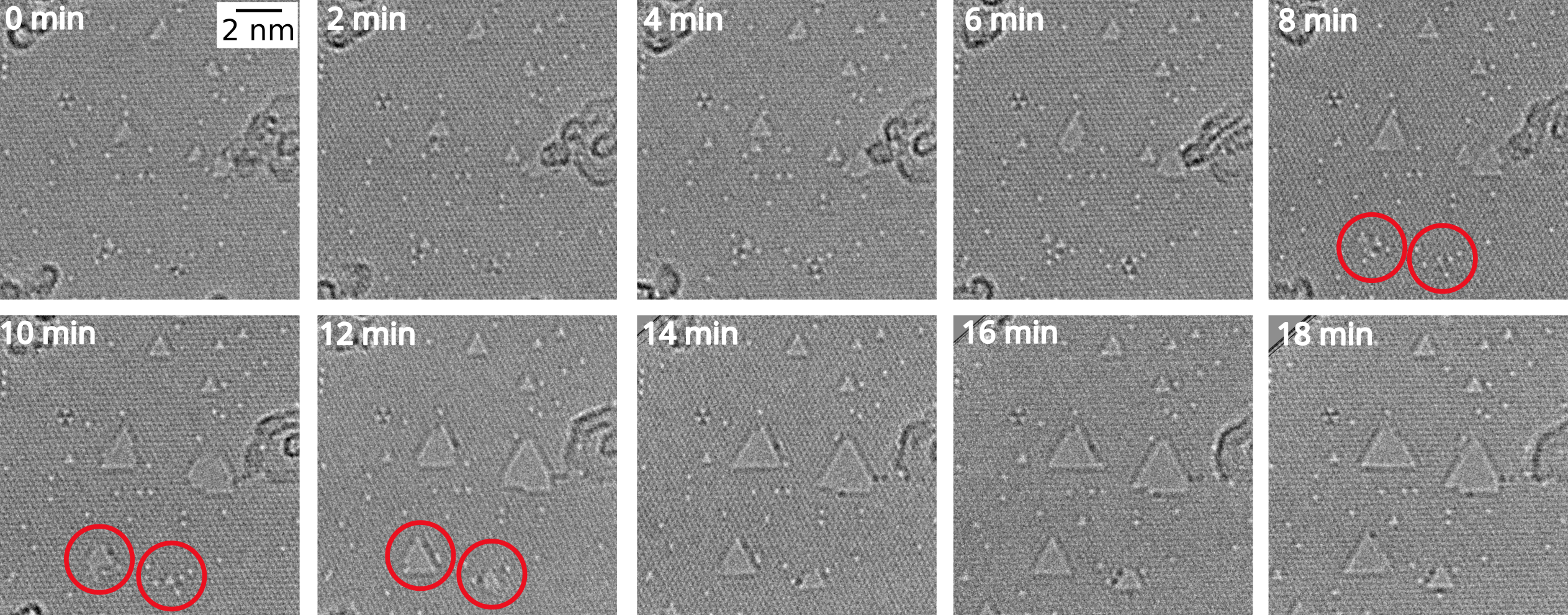}
 \caption{Image series shown in Fig.~4 with circles highlighting two clusters of single atom vacancies that merge to form larger nanopores.}
 \label{Fig:S2}
\end{figure}

\begin{figure}[H]
 \centering
 \includegraphics[width=0.8\textwidth]{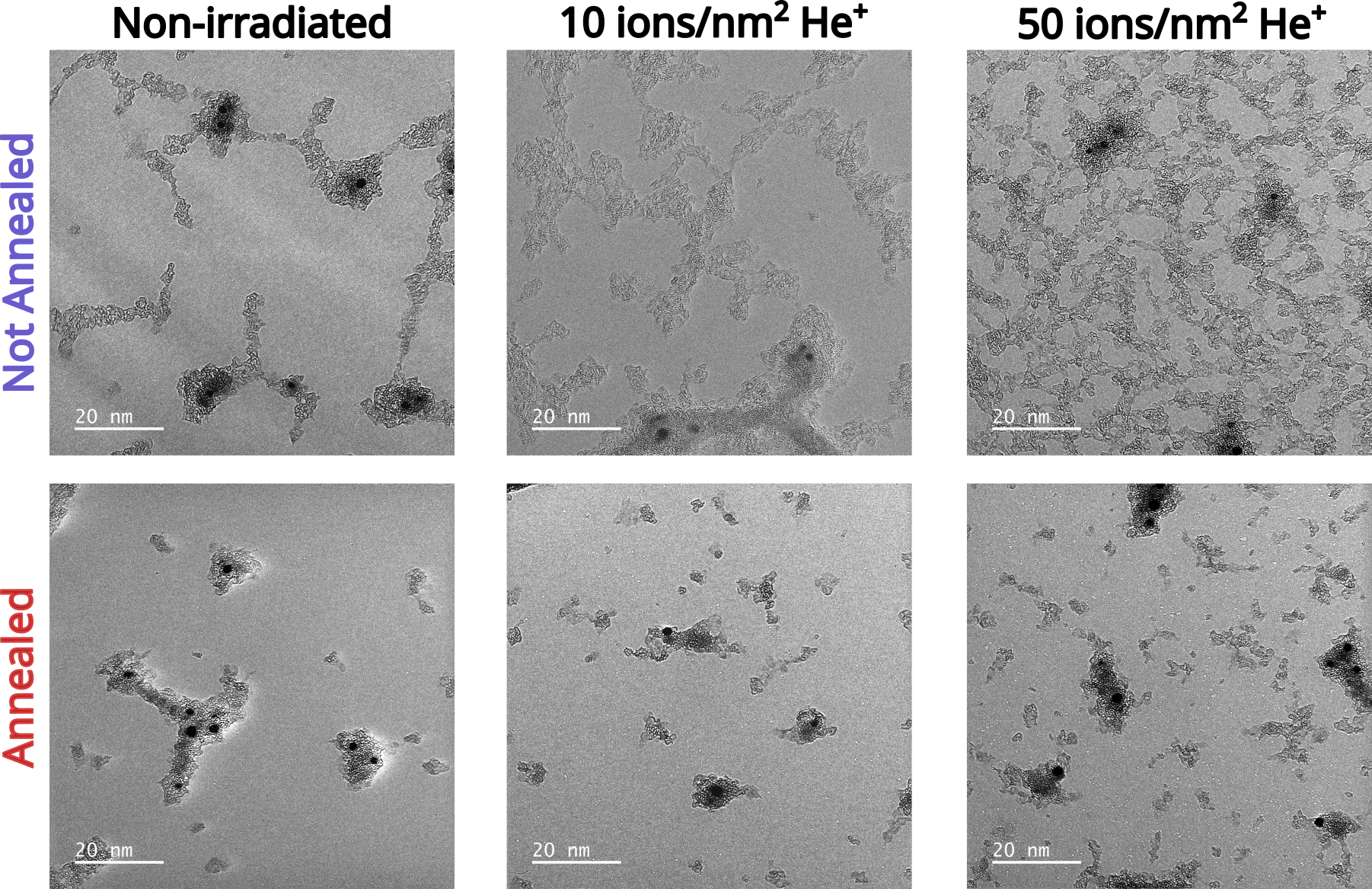}
 \caption{Low magnification TEM images of representative ion-irradiated monolayer hBN with and without annealing treatments.}
 \label{Fig:S3}
\end{figure}

\begin{figure}[H]
 \centering
 \includegraphics[width=0.9\textwidth]{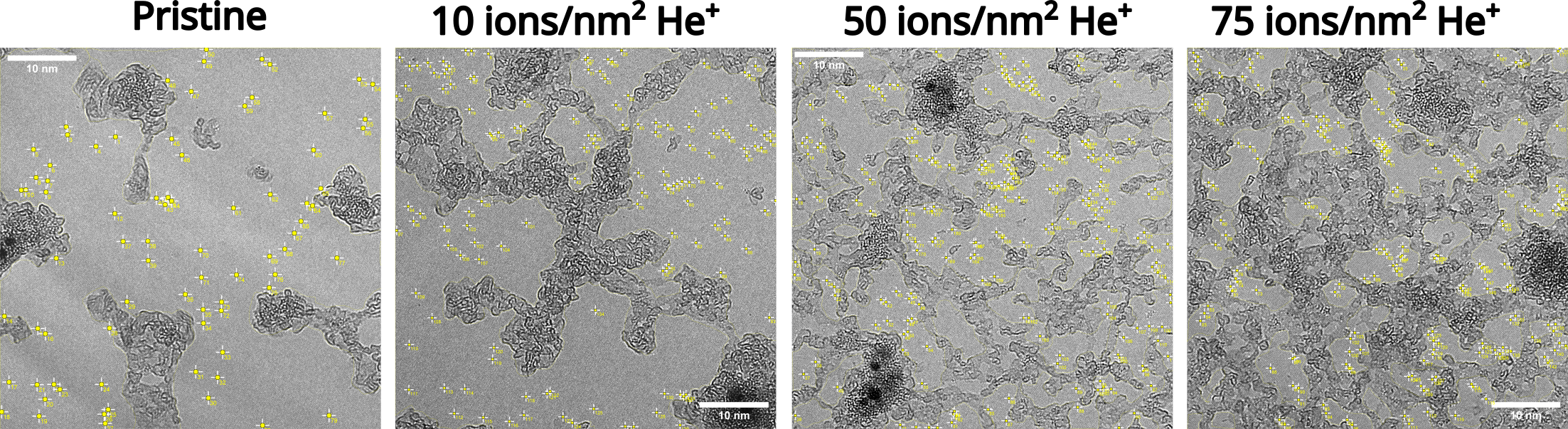}
 \caption{Example TEM images of monolayer hBN with various ion irradiation treatments. Defects were labelled using a numbered multi-point tool.}
 \label{Fig:S4}
\end{figure}

\end{document}